\documentclass[aps,prl,twocolumn,showpacs,preprintnumbers,superscriptaddress,dblfloatfix]{revtex4-1}
\usepackage{graphicx}   % needed for figures
\usepackage{dcolumn}    % needed for some tables
\usepackage{bm}         % for math
\usepackage{amssymb}    % for math
\usepackage{setspace}
\usepackage{amsmath, amssymb, setspace}
\usepackage{array}
\usepackage{booktabs}
\usepackage{mathrsfs}
\usepackage{indentfirst}
\usepackage{slashed}
\usepackage{float}
\usepackage{lmodern}
\usepackage{hyperref}
\usepackage{xcolor}
\usepackage{multirow}
\usepackage{epstopdf}
\usepackage{ulem}
\usepackage{tabularx}
\usepackage[justification=raggedright]{caption}
\usepackage[flushleft]{threeparttable}
\usepackage{hyperref}
\usepackage{soul}

\begin{document}

\preprint
{IPMU16-0168, UCI-TR-2017-02}

\title{Dark Cosmic Rays}

\author{Ping-Kai Hu}
\email[]{pingkai.hu@physics.ucla.edu}
\affiliation{Department of Physics and Astronomy, University of California, Los Angeles\\
Los Angeles, CA 90095-1547, USA}

\author{Alexander Kusenko}
\email[]{kusenko@ucla.edu}
\affiliation{Department of Physics and Astronomy, University of California, Los Angeles\\
Los Angeles, CA 90095-1547, USA}
\affiliation{Kavli Institute for the Physics and Mathematics of the Universe (WPI), UTIAS\\
The University of Tokyo, Kashiwa, Chiba 277-8583, Japan}

\author{Volodymyr Takhistov}
\email[]{vtakhist@physics.ucla.edu}
\affiliation{Department of Physics and Astronomy, University of California, Los Angeles\\
Los Angeles, CA 90095-1547, USA}
\affiliation{Department of Physics and Astronomy, University of California, Irvine\\
Irvine, CA 92697-4575, USA}

\date{\today}

\begin{abstract}
If dark matter particles have an electric charge, as in models of millicharged dark matter, such particles should be accelerated in the same astrophysical accelerators that produce ordinary cosmic rays, and their spectra should have a predictable rigidity dependence.  Depending on the charge, the resulting ``dark cosmic rays" can be detected as muon-like or neutrino-like events in Super-Kamiokande, IceCube, and other detectors.  We present new limits and propose several new analyses, in particular, for the Super-Kamiokande experiment, which can probe a previously unexplored portion of the millicharged dark matter parameter space.  Most of our results are fairly general  and  apply to a broad class of dark matter models.  
\end{abstract}


\maketitle

A preponderance of astrophysical evidence confirms that most of the matter in the universe is dark matter (DM)  that  is not made of ordinary atoms~\cite{Bertone:2004pz}. Little is known, however, about possible non-gravitational interactions of dark matter, which hold the key to identifying  its nature  \cite{Feng:2010gw}.  Most  of the  dark matter candidates need some kind of non-gravitational interaction for their production in the early universe.  Such additional interactions may occur between the DM particles and the Standard Model particles or between  the  DM particles and other components of the dark sector.  In a broad class of well-motivated models the DM particles couple to the Standard Model fields via kinetic mixing~\cite{Holdom:1985ag} and carry a fractional electric charge. 

If DM particles have  an electric charge, they can be accelerated in the same astrophysical environments that generate cosmic rays.  Most  of the  acceleration mechanisms depend not on the charge or mass of the particle, but on  the  rigidity, and, aside from the particulars of injection, one can expect that the ordinary cosmic rays are accompanied by a predictable flux of dark cosmic rays. In this \textit{Letter} we estimate the flux of  dark cosmic rays, explore  detectability of the accelerated particles and present new limits on millicharged dark matter (mDM). 

Models of millicharged dark matter~\cite{Ackerman:mha,Feldman:2007wj,Feng:2009mn,Petraki:2011mv,vonHarling:2012yn} often invoke two or more particles in the dark sector, which may interact by means of an additional $\mathrm{U}(1)_\mathrm{X}$ gauge symmetry associated with a  dark photon.  The dark sector particles  may form dark atoms, or they  may exist in the form of ionized gas~\cite{Goldberg:1986nk,Kaplan:2009de,Cline:2012is,CyrRacine:2012fz,Pearce:2013ola,Petraki:2014uza}.  The production of this form of dark matter in the early universe may rely on asymmetries similar  to  the  matter-antimatter asymmetries~\cite{Petraki:2013wwa,Zurek:2013wia}. 

Due to a kinetic mixing $\tilde{\varepsilon} F'_{\mu\nu} F^{\mu\nu}$ \cite{Holdom:1985ag,Foot:1991kb} between $ \mathrm{U}(1)_{\mathrm{X}} $ and the hypercharge $ \mathrm{U}(1)_{\mathrm{Y}} $ of the Standard Model, dark fermions obtain effective charges and couple to the standard photon.  The kinetic mixing can be eliminated by a field redefinition giving the dark fermions a small electric charge, while the interactions in the dark sector are mediated by a dark photon.  We denote the mass of the resulting dark ions as $m_\mathrm{X}$.  

While models that contain a massive dark photon are phenomenologically rich \cite{Essig:2013lka}, they are already strongly constrained and are also disfavored from more general theoretical arguments \cite{Shiu:2013wxa,Banks:2010zn}. On the other hand, if the dark photon is massless, as we assume in this work,  there exists a largely unexplored region in the $m_\mathrm{X}$ vs. $\varepsilon$ parameter space for $1 \lesssim m_\mathrm{X} \lesssim 100$ GeV \cite{Vinyoles:2015khy}. We will show that it is possible to explore this region with the current experiments. We note that dark electromagnetism can in principle also have an effect on the ion acceleration, but we shall not discuss this possibility here.

It is widely believed that the first order Fermi acceleration (diffusive-shock acceleration) \cite{Blandford:1978ky} is responsible for generating the standard cosmic rays and their power law energy spectra, with supernova (SN) remnants comprising the most probable sources within the Galaxy. Here, charged particles are injected with energies above thermal and get accelerated with each successive pass through a shock wave, generated by the supernova's explosion.  The same mechanism can also accelerate DM ions~\cite{Chuzhoy:2008zy}. One can expect some degree of ionization in atomic dark matter due to an incomplete recombination of the primordial dark-matter gas~\cite{CyrRacine:2012fz}  or due to a later reionization by starlight, supernova explosions~\cite{Foot:2014mia}, and the high-redshift sources that are responsible for reionization of ordinary hydrogen (such as dwarf galaxies, quasars, Population III stars, X-ray sources in the Galactic Center and the galaxy clusters).
For a dark coupling constant $\alpha_D \sim 10^{-2}$
and a mass range of $10 \lesssim m_X \lesssim 1000$ GeV
a $\sim1\%$ global fraction of uniformly ionized DM is allowed \cite{CyrRacine:2012fz}. We stress, however, that these constraints are model dependent.
For our analysis we only assume that DM is fully ionized  locally in the vicinity of acceleration, while outside of this region it can stay neutral and thus the global ionized fraction can be kept very low.
We have estimated that for a range of relevant DM parameters, some of the above-mentioned sources can efficiently overcome the binding energy of dark atoms and produce DM ions.

Fermi acceleration can accelerate particles of electric charge $\varepsilon e$ to the maximum energy \cite{Lagage:1983zz,Hillas:1985is,berezhko1996maximum} 
\begin{equation}
E_{\text{max}}~\sim~\varepsilon e B U L~,
\end{equation}
where $U$ is the shock wave speed, $L$ is the total acceleration length and $B$ is the magnetic field.  For supernova remnants, the relevant length is the size of the shock at the end of the free expansion, $ L \sim 3 $ pc. With a magnetic field of $ B \sim 0.5 \: \mathrm{m} \mathrm{G} $  \cite{Volk:2004vi,Reynolds:2011nk} and $U \sim 0.1$, protons can be boosted to the energy levels above PeV, with the corresponding value for mDM lower by a factor of $\varepsilon$. We have confirmed that energy losses due to synchrotron radiation can in general be neglected here.

While there is a degree of uncertainty in predicting precisely the cosmic ray flux, especially due to the unknown injection spectrum, it is possible to obtain a robust estimate for DM flux
by using results for ion acceleration in shocks. Hence, the DM flux is related to proton flux at equal particle rigidity as
\cite{Malkov:1998in, Malkov:2011gb} (see Ref.~\cite{Ellison:1997an,Berezhko1999}
for an alternative acceleration treatment)
\begin{equation}
\dfrac{dN_X}{dR} \bigg/ \dfrac{dN_p}{dR}
\simeq
\dfrac{(\rho_\mathrm{X} / m_\mathrm{X})}{(\rho_\mathrm{p} / m_\mathrm{p})}
\times \dfrac{e_{\mathrm{inj}}^X}{e_{\mathrm{inj}}^p}~,
\end{equation}
where $ R = p / Q$ is the rigidity of particles with charge $Q$ and momentum $p$.
The enrichment factor $ (e_{\mathrm{inj}}^X / e_{\mathrm{inj}}^p) $ describes the difference between mDM and protons in the shock injection.
Here, we have made the standard assumption of strong shocks and that enrichment due to injection saturates around 4, as expected to occur above $(m_{\mathrm{X}}/m_p)/\varepsilon \gtrsim 3.5$ \cite{Malkov:1998in}.
Assuming Navarro-Frenk-White (NFW) DM profile \cite{Navarro:1995iw} for the DM distribution and a reference point of 1 kpc from the Galactic Center, where a high star formation rate and thus a high SN rate is expected, the dark matter density is $\rho_\mathrm{X} = 4.1$ GeV cm$^{-3}$. The proton number density in the interstellar medium is $ (\rho_p / m_p) = 1 $ cm$^{-3}$.
Hence, the energy spectrum of dark cosmic ray flux 
is predicted to be 
\begin{align} \label{eq:mdmflux}
\dfrac{d N_{\mathrm{X}}}{d E}~&\simeq
\dfrac{(\rho_\mathrm{X} / m_\mathrm{X})}{(\rho_\mathrm{p} / m_\mathrm{p})} \,
\dfrac{e_{\mathrm{inj}}^X}{e_{\mathrm{inj}}^p}
\,\, \varepsilon^{(\alpha -1)} \,
\dfrac{d N_p}{d E} \\[3pt]
~&= 30 \, \varepsilon^{(\alpha -1)} \, \Big( \dfrac{\text{GeV}}{m_\mathrm{X}}\Big) \Big(\dfrac{E}{\text{GeV}}\Big)^{-\alpha}~
\big/ (\mathrm{GeV} \: \mathrm{cm}^2 \: \mathrm{s} \: \mathrm{sr})
,
\notag
\end{align}
% * <pingkai.hu@gmail.com> 2016-12-28T05:21:44.098Z:
%
% ^.
where $(d N_p / d E) \propto E^{-\alpha} $ with $ \alpha = 2.7 $ is the experimentally observed proton flux \cite{Agashe:2014kda}.

Although the standard cosmic rays are highly isotropic~\cite{Agashe:2014kda}, the dark cosmic rays have a larger gyroradius in the same galactic magnetic fields.  When the gyroradius is comparable to or greater than the thickness of the Galactic disk ($\sim 300 $ pc), the arrival directions of dark cosmic rays should exhibit an anisotropy in the direction of the Galactic center. 
For protons, no anisotropy is expected below $ 10^{18} $ eV, but, for mDM, the anisotropy should be observed at energies $\sim \varepsilon\times 10^{18}$~eV.  If detected, such dark cosmic rays should point back to the source. 
A detailed discussion of anisotropies, as well as the acceleration mechanisms different from the Fermi shock acceleration, will be presented in the upcoming publication~\cite{darkrays}. 

\begin{figure}
\centering
\hspace{-2em}
\includegraphics[width=1\linewidth]{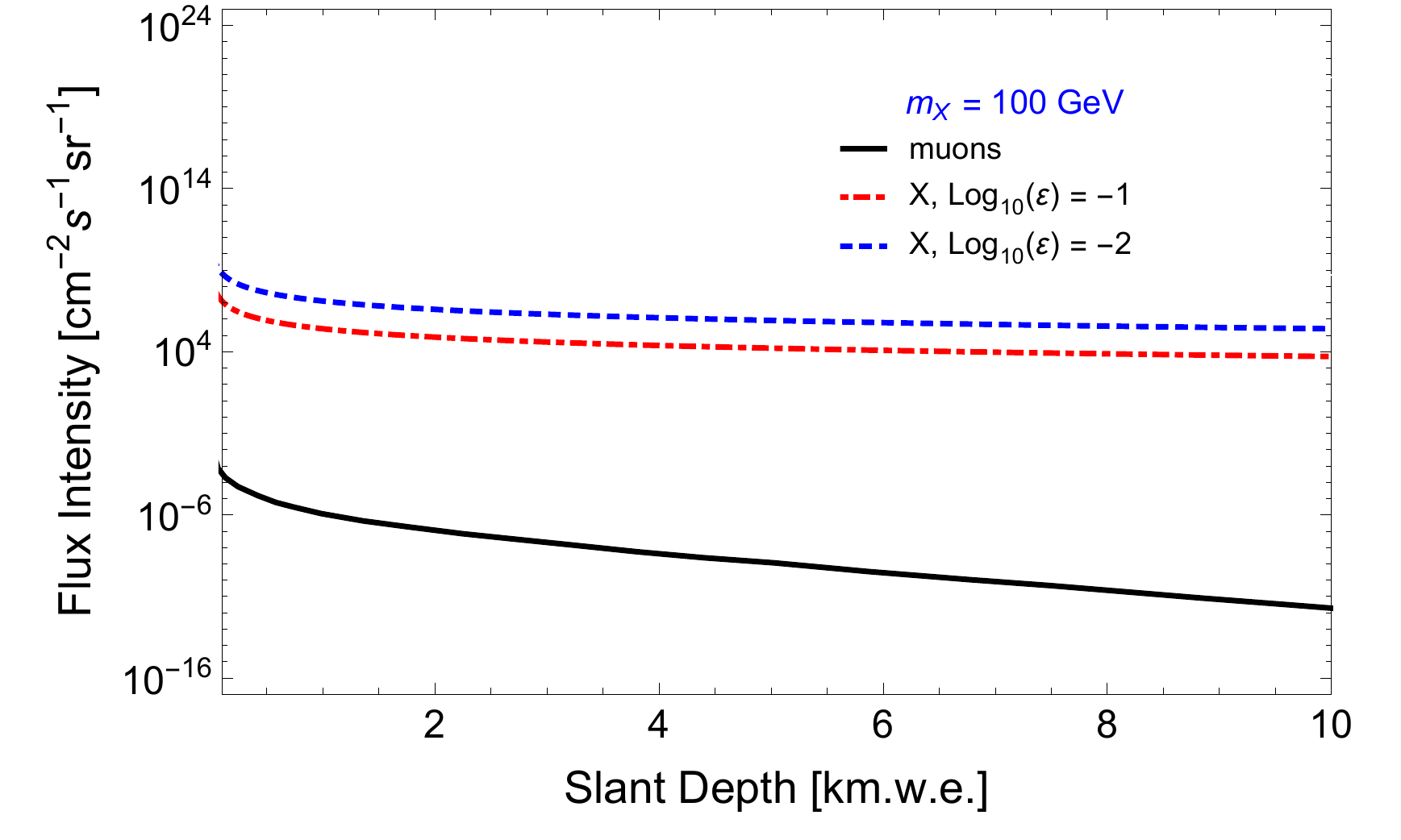}
\caption{(color online) Integrated vertical flux intensity for mDM with $m_\mathrm{X} = 100$ GeV, $\varepsilon = 10^{-0.5}$ (dashed line) and $\varepsilon = 10^{-1}$ (dot-dashed line). Comparison with muon intensity (solid line) from the Crouch curve \cite{Reichenbacher:2007dm} is shown.
}
\vspace{-2em}
\label{fig:intvsdepth}
\end{figure}

\begin{figure*}[htb]
\begin{minipage}[b]{0.45\textwidth}
\centering
\includegraphics[width=\textwidth]{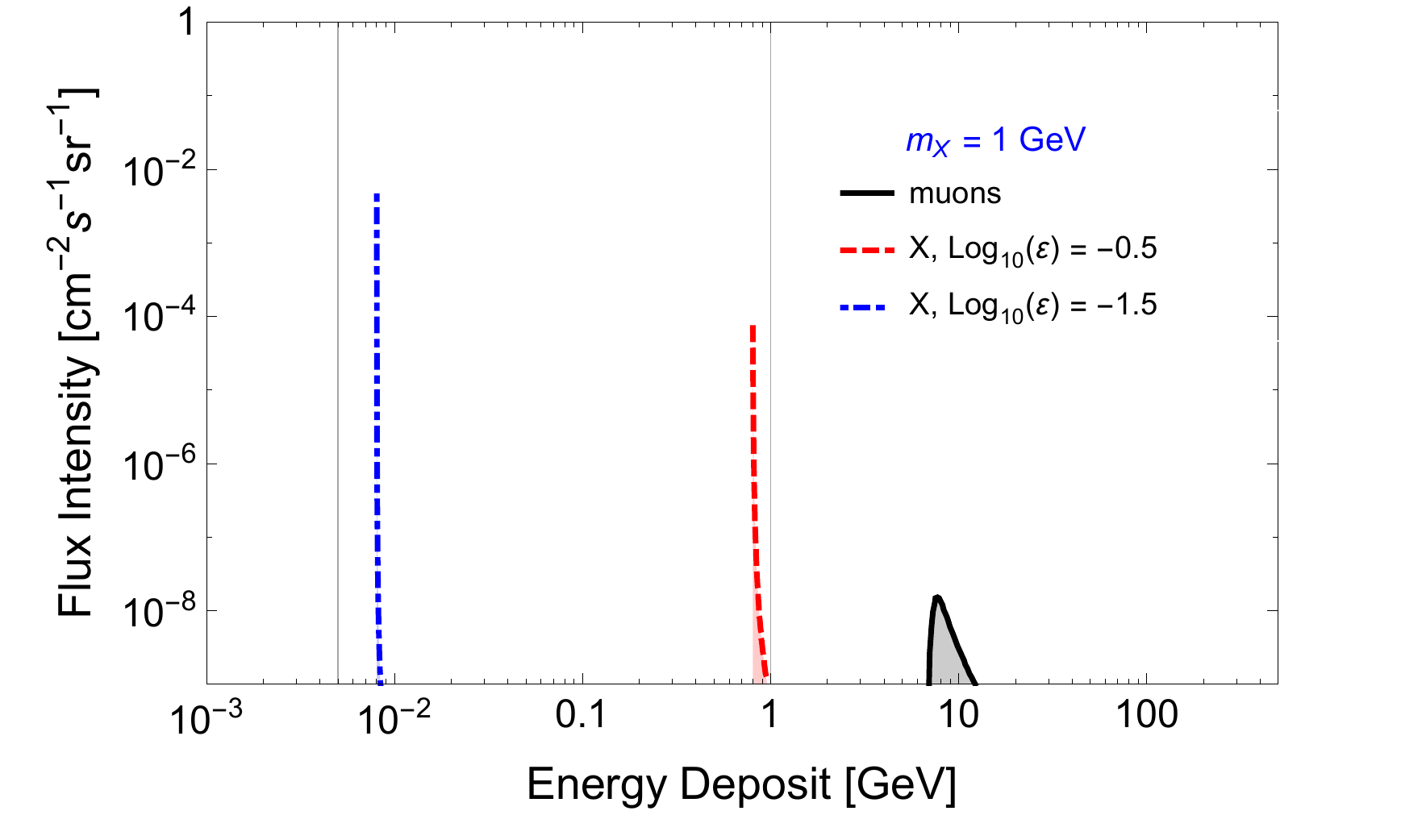}
\end{minipage}
\hspace{3em}
\begin{minipage}[b]{0.45\textwidth}
\centering
\includegraphics[width=1.15\linewidth]{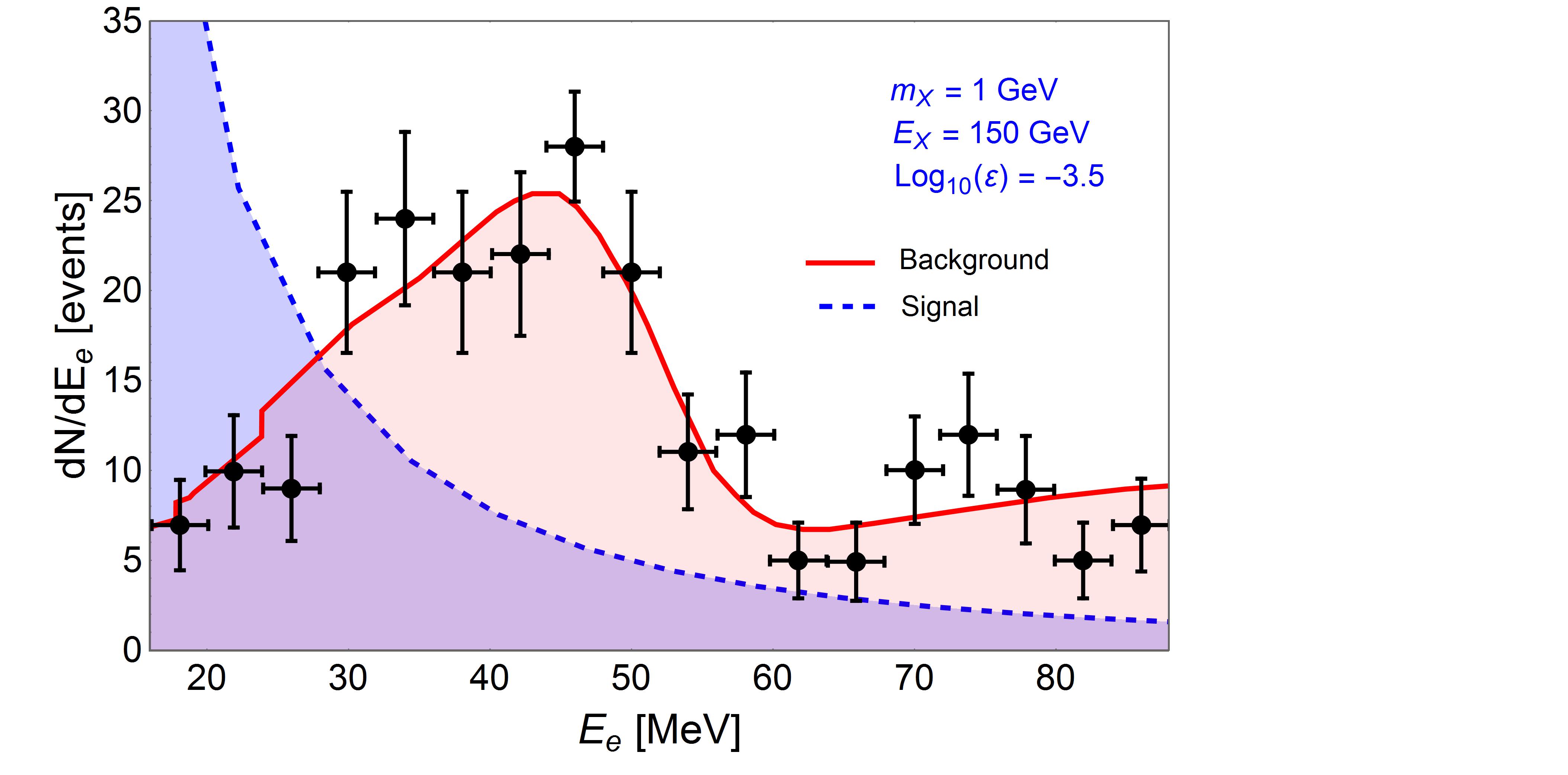}
  \vspace{-0.55em}
\end{minipage}
\caption{(color online) [left] Energy deposit from a vertical through-going flux of mDM ($\mathrm{X}$) in SK, compared with muons from Ref.~\cite{Li:2014sea}.
Results for $m_X = 1$ GeV and $\varepsilon = 10^{-0.5}$ (dashed), $\varepsilon = 10^{-1.5}$ (dot-dashed) are shown. Vertical lines indicate the current SK muon 
 fitter sensitivity of $\sim 1$ GeV as well as the potential fitter improvement down to $\sim 5$ MeV. [right] Recoil electron spectrum from mDM (blue) with $m_\mathrm{X} = 1$ GeV, $\varepsilon = 10^{-3.5}$ and energy of $E_{\mathrm{X}} = 150$ GeV, compared with the background (red) and data (black) of the full SK-I supernova neutrino analysis sample from Ref.~\cite{Bays:2011si}. The peak of around $\sim40$ MeV is from $\nu_{\mu}$ decay electrons, while the rising spectrum is from $\nu_e, \overline{\nu_e}$ interactions. Event rate
is normalized per year.}
\label{fig:detect}
\end{figure*}

Depending on the value of $\varepsilon$,
mDM exhibits either a muon-like or a massive neutrino-like behavior in a detector.~Cherenkov detectors can thus be used to search for the corresponding signals. However, there are important differences between DM and neutrinos that can be used to distinguish between the neutrinos and the dark cosmic rays. Unlike neutrinos, dark cosmic rays will not exhibit oscillations~\cite{Fukuda:1998mi}. Furthermore, they will not show the 6.3 PeV Glashow resonance~\cite{Glashow:1960zz}. The photon-mediated interactions that are relevant for detection are similar to those of boosted dark matter \cite{Agashe:2014yua,Necib:2016aez}, namely, the quasi-elastic (QE) scattering (i.e.~ionization), the photo-nuclear (PN) interactions and the deep inelastic scattering (DIS):
\begin{align}
\text{QE}: & ~~~~~~~~ X + e^{-} \rightarrow X +  e^{-}   \notag\\
\text{PN}: & ~~~~~~~ ~X + N  \rightarrow X + N'  \\
\text{DIS}: & ~~~~~~~ ~X + N  \rightarrow X + N' + \text{hadrons} \notag
\end{align}
Here $N, N'$ are nuclei.~While QE generally provides the most sensitive channel, in the upcoming work~\cite{darkrays} we will demonstrate that DIS interactions can be employed to study deviations in the neutrino flux and also to account for the ultra-high energy neutrino events observed by IceCube~\cite{Aartsen:2014gkd}. 

For $\varepsilon \gtrsim 10^{-2}$ (muon-like) mDM, the Earth is not transparent to the dark cosmic rays and the flux as well as the energy spectrum of the particles that can reach an underground detector will depend on the zenith angle. Following the energy loss calculations for the standard cosmic ray muons~\cite{Agashe:2014kda,Groom:2001kq}, the stopping power for mDM is given by
\begin{equation} \label{eq:dedx}
- \Big\langle \dfrac{dE}{dx} \Big\rangle = a(E) + b(E) E~.
\end{equation}
Here $E$ is the energy, $a(E)$ represents the ionization losses described by the Bethe-Bloch formula and $b(E)$ represents the losses due to radiative processes (bremsstrahlung, pair-production, photo-nuclear effects) that dominate at higher energies:
$b_{\text{total}} = b_{\text{brem}} + b_{\text{pair}} + b_{\text{nucl}}$. The coefficients $a$ and $b$ depend not only on the energy but also on the medium. The point where radiative losses become comparable with ionization losses is defined by the critical energy $\epsilon = a(\epsilon)/b(\epsilon)$. From the leading behavior of the energy loss processes \cite{Groom:2001kq}, one can infer the ratio of 
the coefficients $a_\mathrm{X}(E)$ and $b_\mathrm{X}(E)$ to those for muons, $a_{\mu}(E)$ and $b_{\mu}(E)$:
\begin{align} 
&\dfrac{a_\mathrm{X}}{a_{\mu}} \propto \varepsilon^2; & 
&\dfrac{b_{\mathrm{X}, \text{brems}}}{b_{\mu, \text{brems}}} \propto \Big(\dfrac{m_{\mu}}{m_\mathrm{X}}\Big)^2  \varepsilon^4~;~  \nonumber\\
&\dfrac{b_{\mathrm{X}, \text{pair}}}{b_{\mu, \text{pair}}} \propto \Big(\dfrac{m_{\mu}}{m_\mathrm{X}}\Big) \varepsilon^2; & 
&\dfrac{b_{\mathrm{X}, \text{nucl}}}{b_{\mu, \text{nucl}}} \propto \varepsilon^2~.
\end{align}
Here $m_\mu$ is the muon mass. 

We have confirmed numerically that $e^+e^-$ pair production and bremsstrahlung dominate over the photonuclear contributions. In the parameter space region of interest $m_{\mathrm{X}} > m_{\mu}$, and thus pair production provides the leading behavior, resulting in
\begin{equation}  \label{eq:brel}
\dfrac{b_{X, \text{total}}}{b_{\mu, \text{total}}} \simeq \dfrac{1}{2} \Big(\dfrac{m_{\mu}}{m_X}\Big) \varepsilon^2 ~~~;~~~ \dfrac{\epsilon_X}{\epsilon_{\mu}} \simeq 2 \Big(\dfrac{m_X}{m_{\mu}}\Big)~.
\end{equation}
Since $a, b$ vary slowly with energy, to a good approximation they can be taken as constant. 

The above allows for a simple estimate \cite{Agashe:2014kda} of the energy spectrum after passage of $x$ meter water equivalent (m.w.e) depth of the material: 
\begin{equation} \label{eq:eloss}
E_X (x) = (E_{\mathrm{X}, 0} + \epsilon_\mathrm{X}) e^{-b_\mathrm{X} x} - \epsilon_\mathrm{X}~,
\end{equation}
where $E_{\mathrm{X},0}$ is the initial energy of $\mathrm{X}$. Similarly, for a flux of the form $K E^{-\alpha}$ the integrated vertical flux intensity is given by
\begin{equation}
I_\mathrm{X} (x) = \dfrac{K \epsilon_\mathrm{X}^{- \alpha + 1}}{\alpha-1} e^{-(\alpha-1) b_\mathrm{X}  x} (1 - e^{- b_\mathrm{X} x})^{-\alpha + 1} ~,
\end{equation}
with $\alpha, K$ determined by Eq.~\eqref{eq:mdmflux}. Since the atmospheric density is low, we can focus  without loss of generality only on the flux modulation due to rock propagation.
In Fig.~\ref{fig:intvsdepth} we display the integrated vertical flux intensity vs. depth for several sample parameter space points and compare with the standard approximation for the muon vertical flux intensity as determined by the ``Crouch curve'' \cite{Reichenbacher:2007dm}. Here we have used Eq.~\eqref{eq:brel} as well as $\epsilon_{\mu} = 600$ MeV and $b_{\mu} = 4 \times 10^{-6}$ GeV g$^{-1}$ cm$^{2}$ parameter values for propagation of energetic muons within the standard rock~\cite{Reichenbacher:2007dm}.

Due to a broad MeV - 10 TeV energy reach, a large size and nearly 20 years of collected data (SK-I to SK-IV phases), Super-Kamiokande (SK) \cite{Fukuda:2002uc} can provide the best sensitivity for mDM and will thus be our focus. The 50 kiloton water Cherenkov detector is located at a depth of 2.7 km.w.e. The cylindrical fiducial volume that is used for physics analyses comprises 22.5 kiloton, with a width of 30.3 m and a height of 32.4 m.

Following the standard discussion for charged particles~\cite{Agashe:2014kda},
the Cherenkov radiation \cite{Frank:1937fk} spectrum of mDM signal (number of photons $dN_X$ emitted from $X$ per length of path $dx$ and per unit wavelength $d\lambda$) is given by
\begin{equation} \label{eq:cspec}
\dfrac{d^2 N_X}{dx d\lambda} = \dfrac{2 \pi \alpha_f \varepsilon^2}{\lambda^2} \Big(1 - \dfrac{1}{n^2 \beta^2}\Big)
= \dfrac{2 \pi \varepsilon^2 \alpha_f}{\lambda^2} \sin^2 \theta_C~,
\end{equation}
where $\alpha_f$ is the fine structure constant, $\lambda$ is the emitted light wavelength and $\theta_C$ is the Cherenkov opening angle. Thus, only the normalization but not the distribution of the spectrum changes for mDM.
The particles will only emit light if they are above the Cherenkov energy/momentum threshold of the medium, which for water are $E_{\text{th}} = 1.52 \, m_{\mathrm{X}}$ and $p_{\text{th}} = 1.14 \, m_{\mathrm{X}}$, respectively.

As mDM particles traverse the detector, they deposit energy that can be compared with muons. We assume for simplicity only a vertical mDM flux and that particles fully penetrate the detector. Using the flux of Eq.~\eqref{eq:mdmflux} as an input for Eq.~\eqref{eq:eloss}, we propagate mDM through 2.7 km.w.e. depth of the standard rock to the top of the SK detector and then through 32.4 m.w.e. of water (within the detector itself) to the bottom. To ensure visible signal, we require that the energy is above threshold. The energy difference between the top and bottom of the detector is the energy deposited.
The values of the standard rock and water propagation parameters $b, \epsilon$ \cite{Groom:2001kq} are chosen for muons at the energy of few hundred MeV, which  comprise majority of the flux at Super-K according to simulations \cite{Tang:2006uu}. We show in Fig.~\ref{fig:detect} the energy deposited by mDM particles, as well as the current muon fitter sensitivity that is conservatively estimated at $\sim 1$~GeV, and the potential fitter improvement down to $\sim 5$~MeV, made possible by the newly developed trigger \cite{CARMINATI2015666}.
Probability distribution for muon energy deposition was estimated in Ref.~\cite{Li:2014sea}, which we normalized according to the Crouch curve distribution at the detector's depth. We have confirmed that for muons our analysis gives results in agreement with Ref.~\cite{Li:2014sea}. The observed spread in the muon distribution
comes from detailed  simulations taking into account the energy dependence of $a$ and $b$  as well as energy fluctuations~\cite{Agashe:2014kda,Groom:2001kq}, 
which follow Landau distribution~\cite{Landau:1944if} and lead to a spread tail at higher energies.
We emphasize that this search is nearly background free due to the event characteristics and even a few particles per year can be detected. Since the muon rate at SK is $\sim 2$~Hz~\cite{Habig:2001ei,Tang:2006uu}, only a weak upper bound on $m_{\mathrm{X}}$ sensitivity can be placed.
We highlight that SK can thus be sensitive to $\varepsilon \sim 10^{-2}$ and this result is independent of the flux details, which determine the normalization, and can be competitive with previous dedicated fractional-charge particle searches~\cite{Ambrosio:2000kh,Mori:1990kw}.

\begin{figure}
\centering
\vspace{2em}
\hspace{-2em}
\includegraphics[width=1\linewidth]{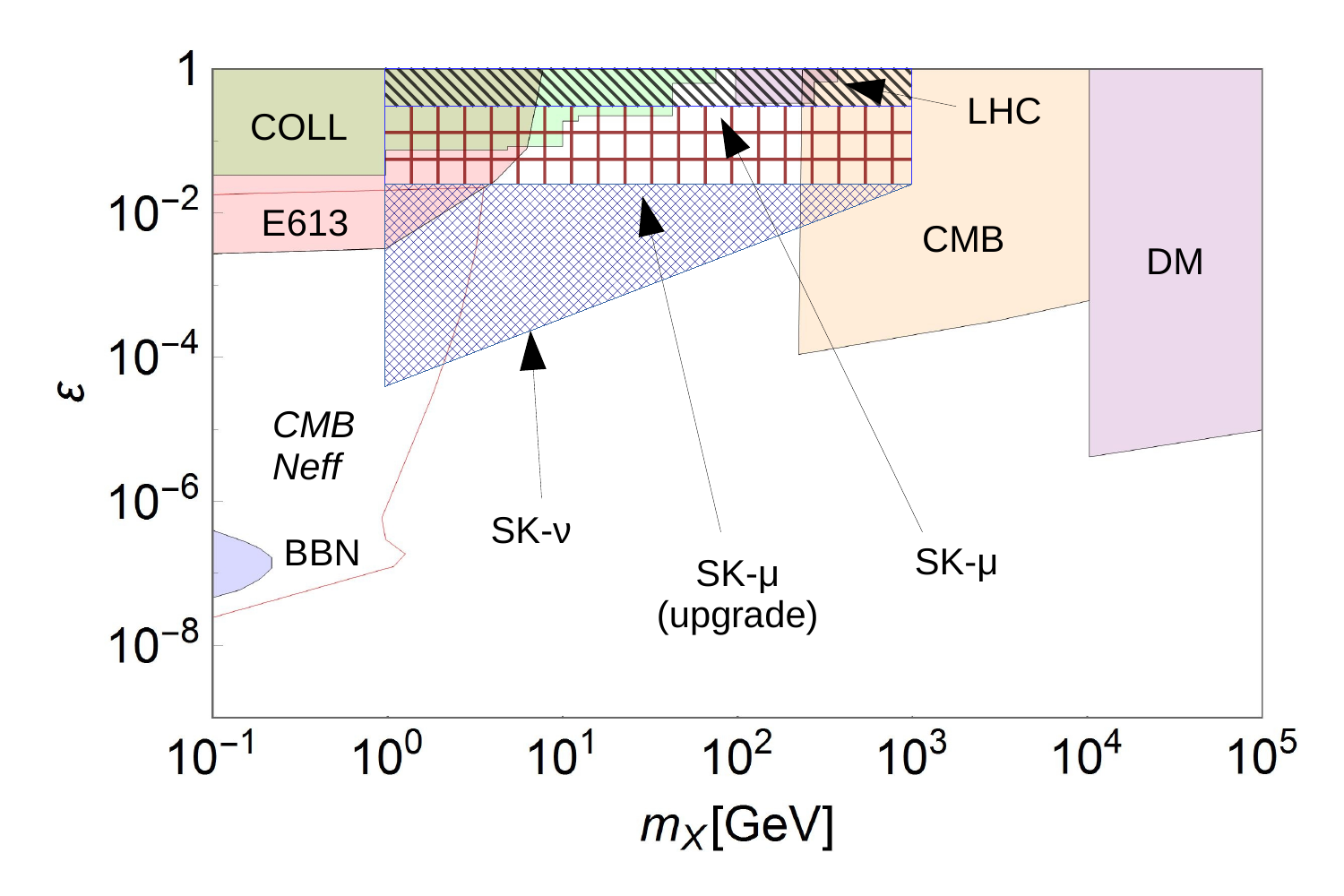}
\vspace{2em}
\caption{(color online) Summary of constraints on millicharged particles in the mass/millicharge plane with a massless dark photon, from Ref.~\cite{Vinyoles:2015khy}. The bounds from CMB and BBN \cite{Vogel:2013raa} using 2015 Planck data \cite{Andreas:2012mt}, collider bounds (COLL) \cite{Redondo:2008aa,Schwarz:2011gu}, dark matter (DM) \cite{Redondo:2008aa}, CMB \cite{Mizumoto:2013jy}, LHC \cite{An:2013yua} and  E613 \cite{Bahre:2013ywa} are shown. Sensitivity reach of Super-K (this work) is shown for muon-like signal (SK-$\mu$, tilted line hatch)  as well as for the recoil electron spectrum (SK-$\nu$, rhombus hatch). Sensitivity reach with possible future muon fitter upgrade is also displayed (SK-$\mu$ upgrade, square hatch). The CMB $N_{\rm eff}$ constraints are, in general, model dependent and can be circumvented, e.g., by decaying sterile neutrinos~\cite{Fuller:2011qy}. 
}
\label{fig:constraints}
\end{figure}

Another mDM signature in SK is single electron-like ring spectrum, resulting from the quasi-elastic interactions. Here we consider a uniform full sky flux. The minimum energy of recoil electron $E_e$ for mDM particle of energy $E_{\mathrm{X}}$ is set by the experimental threshold, while the maximum is set by kinematics. The cross-section is given by 
\begin{equation} \label{eq:eldiffxsec}
\dfrac{d \sigma_{X e^- \rightarrow X e^-}}{d t} =
\dfrac{1}{32 \pi} \dfrac{(\varepsilon e^2)^2}{t^2}
\dfrac{8 E_X^2 m_e^2 + t(t + 2s)}{m_e^2 (E_X^2 - m_X^2)}~,
\end{equation}
where $s = m_X^2 + m_e^2 + 2 E_X m_e$ and $t = q^2 = 2 m_e (m_e - E_e)$. Unlike the boosted dark matter~\cite{Agashe:2014yua}, our cross-section increases with lower transfer momentum $t$ due to a massless mediator. Furthermore, our signal is isotropic and the flux is constant in time. 
An anisotropy can be expected for energies above $\varepsilon\times 10^{18}$~eV, but the predicted flux at these energies is small. 
In SK, the optimal search region for our signal is in the $\sim16-88$ MeV range of the supernovae relic neutrino sample \cite{Malek:2002ns}, since the solar $\sim10-20$ MeV sample \cite{Abe:2016nxk} is plagued by the muon spallation background. For illustration, we display in Fig.~\ref{fig:detect} the recoil electron signal from mDM of energy $E_{\mathrm{X}} = 100$ GeV as well as the $\nu_{\mu}, \nu_{e}$ backgrounds, taken from Ref.~\cite{Bays:2011si}.
To estimate the sensitivity, we perform a simple event counting in the signal region. Here, the average signal detection efficiency is around 90\% \cite{Bays:2011si}. The total yearly event rate in the $E_e^{\text{th}} - E_e^R$ energy region is given by

\begin{align}
\label{eq:fullerecoilrate}
&N_{\text{total}} = C \Delta T \, N_{e, \text{target}} \, \int_{E_X^{\text{min}}}^{E_X^{\text{max}}}  \Big(\Phi_X^{\text{sky}} \times \notag\\ &\times  \int_{E_e^{\text{th}}}^{E_e^{\text{max}}(E_X) \leq E_e^R} \dfrac{d \sigma_{X e^- \rightarrow X e^-}}{d E_e} \, dE_e\Big) dE_X~,
\end{align}
where $C$ is the efficiency, $N_{e, \text{target}}$ is the number of electrons in SK, $\Phi_{\mathrm{X}}^{\text{sky}}$ is the full sky flux, $\Delta T = 1 $ year is the running time and $E_{\mathrm{X}}^{\text{min}}$ is the minimum mDM energy required for the  recoil electron to be within the signal region. If the signal event rate in the region exceeds 1-$\sigma \simeq 5$ events/year error fluctuation in the sample \cite{Bays:2011si}, we denote it as ``observable''. Solving numerically Eq.~\eqref{eq:fullerecoilrate} for a variety of inputs, we obtain a sensitivity in charge of up to $\varepsilon \sim 10^{-4.5}$. Since the background describes the data well, our sensitivity gives an estimate for new limits on mDM. We expect that a comprehensive analysis based on the likelihood method, as in Ref.~\cite{Bays:2011si}, will yield better results with improved error treatment.  

In summary, we have pointed out that, regardless of the origin, dark matter ions can in principle be accelerated by the usual astrophysical mechanisms and the resulting dark cosmic rays could be detectable. For millicharged DM, acceleration is similar to the well-studied acceleration of charged dust, leading to a robust estimate of the flux. 
While we do not expect a significant signal from dark cosmic rays in the current direct detection experiments, large-volume neutrino experiments such as Super-Kamiokande can be highly sensitive to them, and present observations allow us to set new limits on mDM.  The sensitivity estimates summarized in Fig.~\ref{fig:constraints} show that a significant portion of the previously unexplored millicharge DM mass-charge parameter space can already be probed in existing detectors. 
The upcoming Gadolinium upgrade of SK~\cite{Beacom:2003nk,Fernandez:2015vhy} should further improve on these results by a factor of few in $\varepsilon$. Future planned experiments, such as the large water Cherenkov experiment Hyper-Kamiokande \cite{Abe:2011ts, Yano:2016rkf} and the large liquid argon experiment DUNE \cite{Acciarri:2015uup}, will yield further improvements in sensitivity.

\textit{Acknowledgments.}
We thank  D.E. Groom, E.~Kearns, M. Malkov, K. Petraki,  M. Smy, H. Sobel, I. Striganov, C. Walter and R.~Wendell for helpful discussions.
This work was supported, in part, by the U.S. Department of Energy Grant No. DE-SC0009937 and Grant No. DE-SC0009920. A.K. was also supported by the World Premier International Research Center Initiative (WPI), MEXT, Japan. A.K. appreciates the hospitality of the Aspen Center for Physics, which is supported by the National Science Foundation grant PHY-1066293. \(\)

\bibliography{mDM_bib}
\end{document}